\documentclass[aip,rsi,reprint]{revtex4-1} 
\usepackage{graphicx}
\usepackage{color}
\begin{document}

\title{Microwave sidebands for laser cooling \\by direct modulation of a tapered amplifier}

\author{J. Mahnke}
\email[]{Electronic mail: mahnke@iqo.uni-hannover.de}
\affiliation{Institut f\"ur Quantenoptik, Gottfried Wilhelm Leibniz Universit\"at Hannover, Welfengarten 1, 30167~Hannover, Germany}

\author{S. Kulas}
\affiliation{ZARM, Universit\"at Bremen, Am Fallturm, D-28359 Bremen, Germany}

\author{I. Geisel}
\author{S. J\"ollenbeck}
\author{W. Ertmer}
\author{C. Klempt}
\affiliation{Institut f\"ur Quantenoptik, Gottfried Wilhelm Leibniz Universit\"at Hannover, Welfengarten 1, 30167~Hannover, Germany}

\date{\today}

\begin{abstract}
Laser cooling of atoms usually necessitates several laser frequencies. Alkaline atoms, for example, are cooled by two lasers with a frequency difference in the Gigahertz range. This gap cannot be closed with simple shifting techniques. Here, we present a method of generating sidebands at $6.6$~GHz by modulating the current of a tapered amplifier which is seeded by an unmodulated master laser. The sidebands enable trapping of $1.1\times 10^{9}$ $^{87}$Rb atoms in a chip-based magneto-optical trap. Compared to the direct modulation of the master laser, this method allows for an easy implementation, a fast adjustment over a wide frequency range and the simultaneous extraction of unmodulated light for manipulation and detection. The low power consumption, small size and applicability for multiple frequencies benefits a wide range of applications, reaching from atom-based mobile sensors to the laser cooling of molecules.
\end{abstract}

\maketitle


\section{Introduction}
Developments in generation and manipulation of laser-cooled atoms have opened exciting perspectives for the improvement of high-precision sensors. Modern atom interferometry allows for the measurement of rotations~\cite{Leveque2009, Stockton2011, Tackmann2012}, accelerations~\cite{Kasevich1991, McGuinness2012}, gravity~\cite{Louchet-Chauvet2011}, gravity gradients~\cite{Fixler2007, Lamporesi2008} and time~\cite{Wynands2005, Deutsch2010, KleineBuning2011}. There is even discussion proposing its use in detection of gravitational waves~\cite{Tino2007, Dimopoulos2008, Delva2009, Hogan2011, Hohensee2011}. Along with striving for better precision, many applications put strict requirements on the physical size and power consumption of such novel sensors~\cite{Schmidt2011}. These restrictions are particularly fundamental for future space missions with their large variety of scientific objectives ranging from fundamental tests of general relativity to geodetic earth observation. Compactness is also a key requirement for the development of mobile commercial sensors.

High-precision atom interferometers rely on a robust source of laser-cooled ultracold atoms. Typically, such setups employ diode lasers due to their ease of operation, reliability, compactness, low cost and low power consumption. Almost all atomic species demand multiple laser frequencies for efficient cooling and manipulation. For example, the wide-spread alkali atoms require at least two laser frequencies for cooling and repumping light. They address the two hyperfine ground states, which are typically separated by several Gigahertz. Modern laser systems thus consist of two frequency-stabilized external-cavity diode lasers, which are typically amplified by a semiconductor laser amplifier, the tapered amplifier (TA).

In principle, one of the two master lasers can be replaced by modulating the light of the first laser. For modulation frequencies of hundreds of Megahertz, this task can be accomplished by standard optical components like acousto- or electro-optical modulators. However, this method requires an additional component and becomes very inefficient for large frequencies. These disadvantages can be avoided by direct modulation of the laser current~\cite{Ezekiel1991,Myatt1993,Stern2010}. By a careful adjustment of the external cavity, this technique has been improved to even suppress the carrier frequency~\cite{Waxman2009}. It has been applied to generate both repumping light for laser cooling as well as phase coherent Raman beams for state manipulation~\cite{Ringot1999}. However, the direct modulation of the master laser suffers from two drawbacks. First, it is difficult to adjust the external cavity of the diode laser for two frequencies. Even more importantly, it is impossible to extract any unmodulated laser light, as needed for manipulation and detection. A promising approach is the direct modulation of a seeded slave laser~\cite{Kowalski2001,Sahagun2012}, although the total available power is still restricted to typically $100$~mW.

In this publication, we demonstrate the generation of sidebands by directly modulating the current of a tapered amplifier. By employing a modulation frequency of $6.6$~GHz, the generated laser light is able to address both the cooling and the repumping transition of $^{87}$Rb. The laser system provides $1$~W of cooling light and enough repumping light to operate a magneto-optical trap with a maximum number of $1.1\times 10^9$~atoms and an initial loading rate of $1.2\times 10^9$~atoms/s. The generation of Gigahertz sidebands by modulation of a TA joins several advantages. It can be realized in a potentially very compact setup with a single master-amplifier combination. The modulation frequency can be freely chosen in a wide range up to several Gigahertz. The master laser's frequency lock is not influenced by the modulation and provides unmodulated laser light. In the future, such a modulation could allow for the generation of phase-coherent Raman beams. Conceptually, this scheme opens an excitingly simple possibility for the generation of a large number of frequencies, as necessary for direct laser cooling of molecules~\cite{Shuman2010}.

\section{Modulation setup}

\begin{figure}
\includegraphics{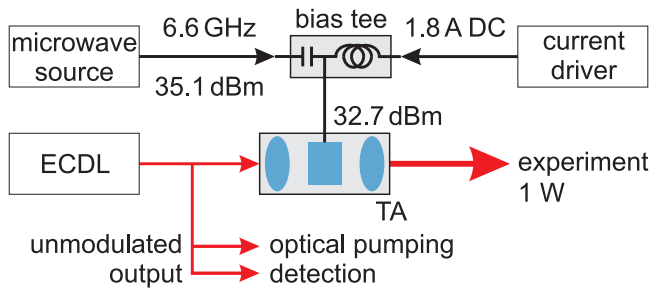}
\caption{\label{fig:microwave} Sketch of the modulation setup. The bias tee combines the microwave signal with the supply current. The modulated current is applied to the tapered amplifier (TA), which amplifies the light from the external-cavity diode laser (ECDL) and generates the desired sidebands on its $1$~W output beam.}
\end{figure}

The generation of sidebands is achieved by directly modulating the supply current of the TA (Eagleyard EYP-TPA-0780-01000-3006-CMT03-0000) with the desired microwave frequency. A sketch of the modulation setup is shown in Fig. \ref{fig:microwave}. The desired microwave frequency of $6.596$~GHz is provided by a phase-locked oscillator (Miteq BCO-130-140-06750-15P), which is seeded by a frequency synthesizer at $132$~MHz. The microwave is amplified by a $3$~W amplifier to $35.1$~dBm. A low-loss circulator is included to limit backreflections. The microwave signal and the DC supply current of $1.8$~A are combined by a bias tee (Pasternack PE1615). It transmits a microwave output power of $32.7$~dBm and is connected to the TA-chip by a $6$~cm long shielded coaxial cable. The connection is not matched to the TA's impedance, therefore more than $99.9\%$ of the power is reflected.

The TA is optically seeded by an external-cavity diode laser, which is frequency-stabilized by saturated absorption spectroscopy to the $F=2 \rightarrow F'=1 / F'=3$ cross-over line of $^{87}$Rb. An acousto-optical modulator shifts the laser frequency to a detuning of $28$~MHz below the $F=2 \rightarrow F'=3$ cooling transition. Part of the light is used for manipulation and detection while the residual $16$~mW are injected into the TA. This saturates the output power of the TA at $1$~W at a constant supply current of $1.8$~A. The output light passes through an optical isolator and is subsequently delivered to the magneto-optical trap.

The microwave-modulated supply current of the TA generates two symmetrical sidebands. These are detected on a high-speed photodiode, where the output of the TA is mixed with an independent, detuned laser beam. This beating technique shifts the sideband signal away from the modulation frequency and is therefore insensitive to parasitic signals by microwave stray fields generated by the high-power microwave source. The resulting power of the beat signal is proportional to the power in the optical sidebands and was used to optimize the entire setup. The beat signal also shows that no microwave signals couple into the master laser. It is difficult to measure the absolute power in the microwave sidebands, as the transfer function of the photodiode is not known. A comparison of the transmission peaks in an additional optical cavity however yields a rough estimate of the relative power $P_{\text{sidebands}}/P_{\text{carrier}}=0.003$ in each sideband. This efficiency is large enough to test the operation of a magneto-optical trap.

\section{Magneto-optical trap setup}
The experimental setup for cooling and trapping $^{87}$Rb in a magneto-optical trap has been described previously~\cite{Jollenbeck2011}. Briefly, it includes a double magneto-optical trap (MOT) consisting of a 2D$^+$-MOT and a 3D-MOT. The 2D$^+$-MOT is operated with coils producing a two-dimensional quadrupole field. The quadrupole field for the 3D-MOT is created by a mesoscopic atom chip. In previous experiments, the light for both traps was produced by two external-cavity diode lasers, one for cooling and one for repumping. Two parallel TAs amplify the combined cooling and repumping light for the two MOTs respectively. One of the 3D-MOT beams is left without repumping light to minimize stray light effects later in the experimental cycle. The operation of a magneto-optical trap with the generated microwave sidebands is demonstrated by replacing all beams of the 3D-MOT which contain repumping light.

\begin{figure*}
\includegraphics{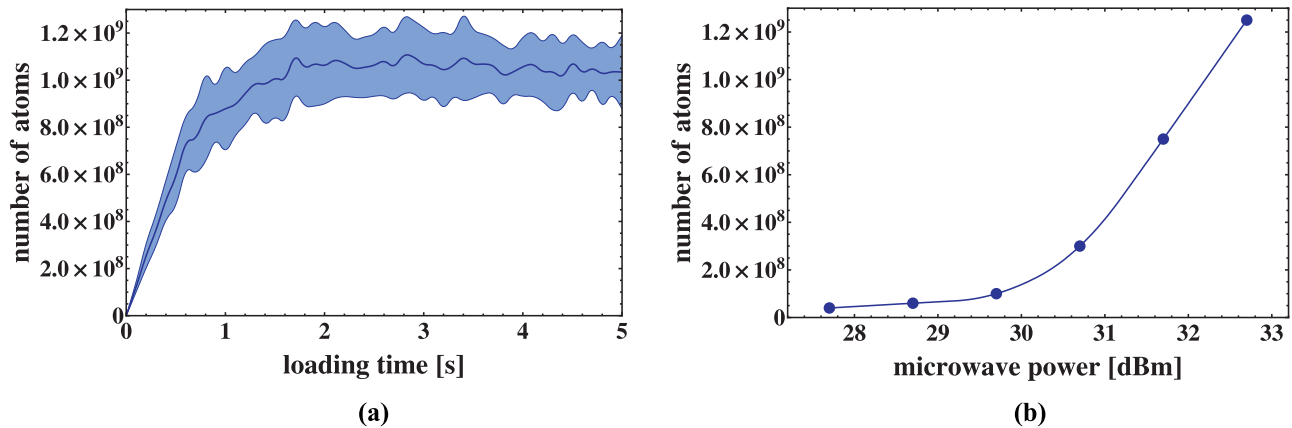}
\caption{\label{fig:loading} (a) Number of atoms in the 3D-MOT as a function of the loading time. The shaded area corresponds to the standard deviation of 6 images. (b) Number of atoms in the 3D-MOT as a function of the total available microwave power behind the bias tee.}
\end{figure*}

\section{Results}

We demonstrate the functionality of our TA current modulation technique by measuring the number of trapped atoms in the 3D-MOT as a function of the loading time (Fig. \ref{fig:loading} (a)). For comparability, the 2D$^+$-MOT and the detection were still operated with the original repumping beam, but the 3D-MOT contained only repumping light that was generated with the modulation technique. The initial loading rate is $1.6\times 10^{9}~\text{atoms}/\text{s}$ reaching a saturated atom number of $1.1\times 10^{9}$ after $1.7$~s. These numbers are about a factor of 10 smaller than those reached with the standard repumping power. However, the successful operation of a MOT with $1.1\times 10^{9}$ atoms shows that he sideband generation is a powerful technique for a wide range of applications.

The size of the MOT is limited by the power in the sidebands, as can be seen in Fig. \ref{fig:loading} (b). It depicts the atom number as a function of the microwave power applied to the TA. The microwave power was reduced by including different attenuators in front of the bias tee. The graph shows no saturation yet, implying that more repumping power should increase the achievable number of atoms. An increase can be accomplished by improving the coupling to the TA with matched impedance or by using a microwave source with higher power. 

The microwave sidebands were also tested in a configuration for a $\sigma^+$-$\sigma^-$ optical molasses~\cite{Chu1986}. A molasses phase with $1.3$~ms duration led to a decreased temperature of $87~\mu$K which compares well to the results of the standard setup.

\begin{figure*}
\includegraphics{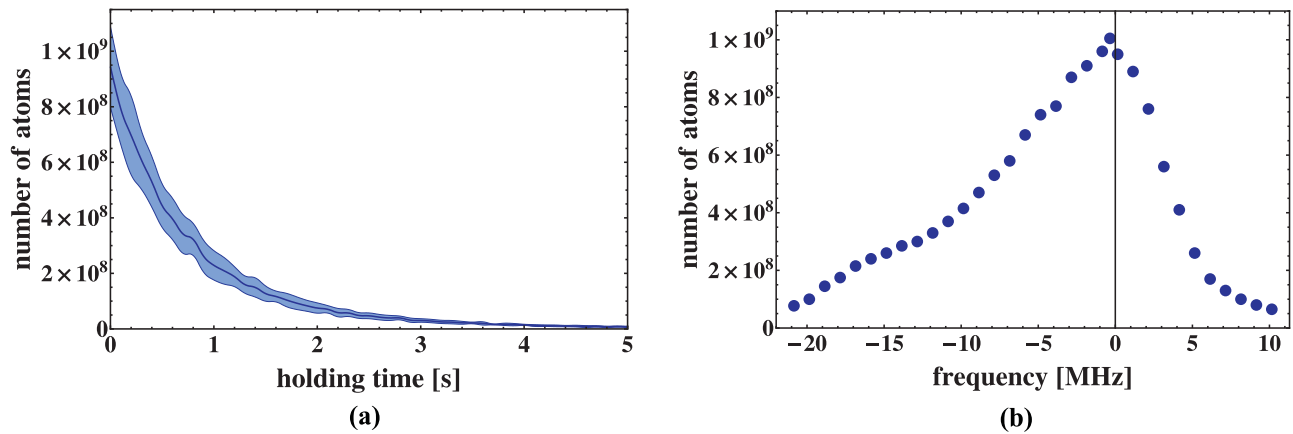}
 \caption{\label{fig:frequency} (a) Number of atoms in the 3D-MOT as a function of the holding time after the 2D$^+$-MOT is switched off. The loading time is $1$~s.	The shaded area corresponds to the standard deviation of 6 images. (b) Number of atoms in the 3D-MOT as a function of the optical sideband detuning with respect to the repumping transition. The solid line indicates the resonance of the $F=1\rightarrow F'=2$ transition.}
\end{figure*}

Another important characteristic is the lifetime of the atoms inside the trap. The lifetime was determined by loading the 3D-MOT, switching off the 2D$^+$-MOT and measuring the exponentially decreasing number of atoms as a function of the holding time (Fig. \ref{fig:frequency} (a)). The resulting lifetime is $0.84$~s, being about half as long as the original lifetime. The reduced repumping power leads to atoms in the $F=1$ state that are not pumped back before they leave the trap.

Finally, the frequency dependence was measured by tuning the initial modulation frequency and measuring the corresponding number of atoms (Fig. \ref{fig:frequency} (b)). It is apparent that the highest atom count is achieved with the repumping light in resonance. The asymmetry reflects that the MOT is enhanced by red-detuned light because it realizes the MOT-configuration also for the repumping states.

\section{Conclusion and Outlook}

The TA modulation is a simple and robust technique to generate multiple laser frequencies in an atom-optics experiment. We have demonstrated that it is possible to load a $^{87}$Rb MOT with $1.1\times 10^{9}$ atoms. A further increase of the power in the sidebands would improve the trappable number of atoms. It should be possible to increase the power by using a better suited TA package or by improving the coupling efficiency with optimized impedance matching. Additionally, the necessary seeding power should decrease by using the TA in a double pass~\cite{Bolpasi2010} configuration.

Many applications can benefit from the creation of such a large number of atoms in a compact apparatus. Furthermore, the method features low power consumption and tuneability within a large frequency range. The generation of multiple sidebands presents a promising outlook for more complex cooling schemes, like the cooling of molecules.

\begin{acknowledgments}
We acknowledge support from the Centre for Quantum Engineering and Space-Time Research (QUEST), from the Hannover School for Laser, Optics and Space-Time Research (HALOSTAR), and from the Research Training Group 1792 "Fundamentals and applications of ultra-cold matter".
\end{acknowledgments}

\end{document}